# Data protection psychology using game theory

Mike Nkongolo[1] and Jahrad Sewnath [2]

University of Pretoria
mike.wankongolo@up.ac.za;
u21442534@tuks.co.za

**Abstract.** With the increasing reliance on technology and the proliferation of personal data, it is crucial to examine how individuals perceive and engage with data protection practices. This research investigates the psychological factors influencing individuals' awareness and understanding of data protection measures. The study utilizes a game theoretical approach, employing strategies, moves, rewards, and observations to gain comprehensive insights concerning psychological factors of data protection. By analyzing the player strategies or moves, several psychological factors impacting data protection awareness are identified, including knowledge levels, attitudes, perceived risks, and individual differences. The findings shed light on the complexity of human cognition and behavior in relation to data protection, informing the development of effective awareness campaigns and educational initiatives.

## 1 Introduction

In the present digital age, as our reliance on technology expands and personal data becomes readily accessible, safeguarding data has emerged as a matter of utmost importance [1]. As individuals increasingly engage in online activities and entrust their sensitive information to various platforms, it is crucial to examine how they perceive and engage with data protection practices [1]. Understanding the psychological factors that influence individuals' awareness and understanding of data protection measures is essential for developing effective strategies to safeguard personal data and promote responsible data management. This paper aims to delve into the intricate dynamics of psychology underlying individuals' awareness and understanding of data protection. By employing a game theoretical approach that combines strategies, moves, and observations, this study seeks to gain comprehensive insights into the psychological factors at play. In his book, Professor Jean-Paul Nkongolo Mukendi defined psychology as a scientific study of the human mind and behavior that explores various aspects of human cognition, emotions, motivations, perception, and social interactions [2]. Moreover, psychology is composed of an affective/emotional system that imprints personal value on environmental data using a cognitive system that collects, processes and interprets information; and finally, a motivational system that triggers and directs the exchange with the environment [2]. Therefore, the psychological factors of data protection refer to the internal processes, beliefs, attitudes, emotions, and cognitive functions that influence human behavior and decision in securing data or information (Fig. 1).

Jean-Paul Nkongolo Mukendi described human behaviors as a set of complex reactions that arise from stimulus and are cleverly orchestrated in the black box [2][1]. The latter imprints on the behavior its singular and differential dimension. In the context of data protection, psychological factors play a crucial role in understanding how individuals perceive, react to, and engage in behaviors related to protecting their data and maintaining privacy. While there is a lack of well-established psychological theories specifically addressing data protection games, the relevance of psychology in this domain can be understood by examining the contributions of renowned psychologists [2].

For instance, Ivan Pavlov's work on classical conditioning and stimulus-response relationships highlights the potential for psychological tests or interventions to influence behavior [2, 3]. Similarly, Alfred Binet's development of the Intelligence Quotient (IQ) test and Edward Thorndike's research on learning and problem-solving demonstrate the use of psychological assessments to measure cognitive abilities and understand human performance [3].

The aim of the proposed data protection game, from a psychological perspective, shares similarities with the psychological tests. While the focus is on data protection rather than traditional intelligence or conditioning, the game is designed to assess specific cognitive functions related to data protection. These functions include risk perception, privacy awareness, threat awareness, security knowledge, and decision-making abilities. The game is updated from Mike Nkongolo [4] and redesigned to enhance individuals' knowledge, attitudes, and understanding of data protection.

### 1.1 Psychological Factors of Data Protection

Protecting data from unauthorized access and misuse involves not only technical measures but also understanding the psychological factors that influence individuals' behaviors and attitudes towards data protection [5]. This section explores various psychological factors associated with data protection as illustrated in Fig. 1. The psychology of data vulnerability encompasses a range of concepts that impact individuals' perceptions and behaviors towards their data.

---

[1] The term black box refers to a psychological model that views the mind as a black box, where inputs (stimuli) are processed and produce outputs (behavior) without explicitly examining the internal mental processes that occur in between.

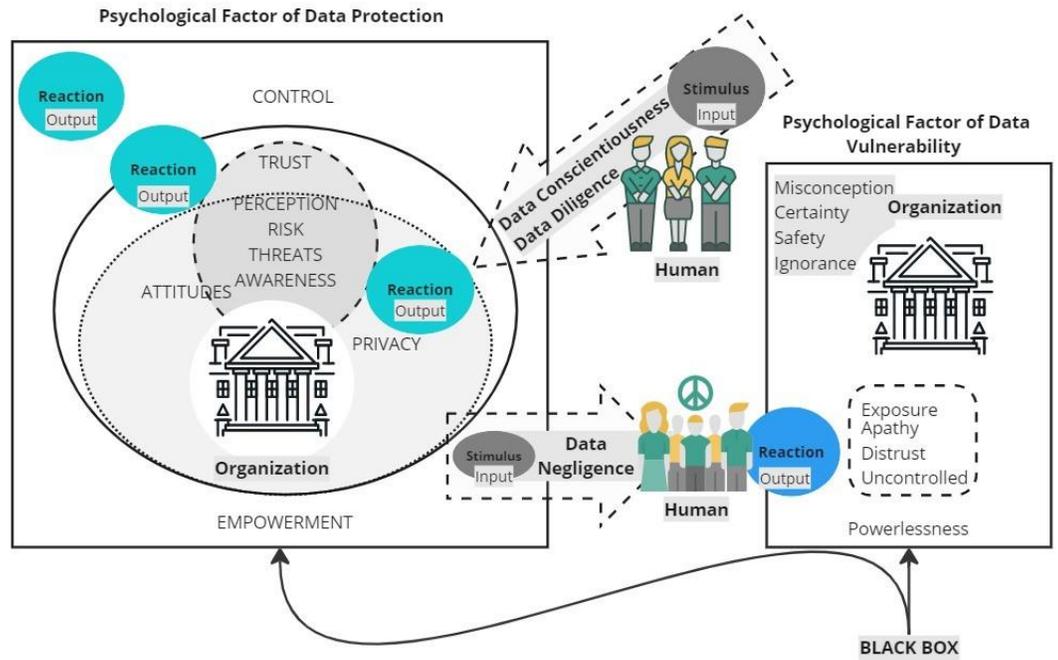

**Fig.1.** The psychological factors of data. Data diligence and negligence are stimulus of data protection and data vulnerability of an organization.

By exploring these concepts, we can gain insights into data vulnerability and its implications. When individuals feel a lack of control over their data, it can generate a sense of vulnerability [6]. This includes concerns about unauthorized access, misuse, or loss of data. Distrust also plays a significant role in data vulnerability. When individuals lack confidence in the security measures, data handling practices, or intentions of data collectors or managers, they perceive an increased risk of data breaches or privacy violations.

Apathy, characterized by a lack of interest or concern, can contribute to data vulnerability [6, 7]. This manifests as a disregard for privacy practices or a lack of motivation to actively protect personal data, making individuals more susceptible to privacy risks and data breaches. Exposure refers to the vulnerability of personal data being accessible to unauthorized parties, which heightens individuals' perceived vulnerability to potential misuse or exploitation [7,8].

Misconception, arising from misunderstandings or incorrect beliefs about data privacy and security, can lead individuals to make uninformed decisions or engage in risky behaviors, inadvertently increasing their vulnerability to data breaches or privacy violations [9]. Safety, on the other hand, reflects individuals' perception of data protection measures and their effectiveness in safeguarding personal information [5, 9]. When individuals believe appropriate safety measures are in place, their perception of vulnerability may decrease.

Ignorance, stemming from a lack of knowledge or awareness about data privacy risks and best practices, leaves individuals more vulnerable to data breaches or privacy violations [10, 11]. The feeling of powerlessness arises when individuals believe they have limited control or influence over the protection of their data. This sense of powerlessness heightens vulnerability and diminishes proactive efforts to protect personal information. Understanding these psychological factors of data vulnerability can inform strategies and interventions aimed at promoting data privacy awareness, encouraging responsible data handling practices, and empowering individuals to effectively protect their personal information.

### 1.2 Risk perception and threat awareness

One crucial psychological factor in data protection is individuals' risk perception and threat awareness. Research has shown that people's perception of risks and threats associated with data breaches or privacy violations influences their behaviors towards protecting their personal information [12]. For example, individuals who perceive higher risks are more likely to engage in protective behaviors and adopt privacy-enhancing measures [5]. Understanding the factors that shape risk perception and threat awareness can provide insights into designing effective interventions for data protection.

### 1.3 Privacy concerns and attitudes

Privacy concerns and attitudes play a significant role in individuals' decision-making regarding data protection. Studies have identified privacy as a fundamental human need and highlighted the impact of privacy concerns on individuals' behaviors in online contexts [5,13]. Positive privacy attitudes are associated with higher intentions to protect personal data and engage in privacy-enhancing behaviors [14]. Examining the antecedents and consequences of privacy concerns and attitudes can help inform strategies to promote data protection.

### 1.4 Trust and trustworthiness

Trust is a crucial psychological factor that influences individuals' willingness to share personal information and engage in data protection practices. Research suggests that individuals are more likely to disclose personal data to entities they perceive as trustworthy [15]. Trust can be influenced by various factors, such as perceived benevolence, competence, integrity, and security [14, 15]. Understanding the dynamics of trust and trustworthiness in the context of data protection can guide the development of trustworthy systems and communication strategies.

### 1.5 User control and empowerment

Providing users with a sense of control and empowerment over their personal data has been found to positively impact data protection behaviors. Studies have shown that individuals who perceive greater control over their personal information are more likely to engage in privacy-protective behaviors [16]. Empowering individuals through user-centric design, transparency, and control mechanisms can enhance their data protection practices and mitigate privacy concerns.

This section has highlighted key psychological factors that influence individuals' behaviors and attitudes towards data protection. Risk perception, privacy concerns, trust, and user control are among the essential psychological factors that shape individuals' decision-making in protecting their personal data [14–16]. Understanding these factors can inform the design of effective interventions, policies, and technologies to promote data protection and privacy.

### 1.6 Research question

Based on the problem statements, this study has formulated the research question as follows:

- How do psychological factors influence individuals' awareness of data protection measures,
- and how can a game theoretical approach provide insights into these factors?

Based on the research question, the research hypothesis can be formulated as follows:

- **Hypothesis 1**: Psychological factors significantly influence individuals' awareness of data protection measures. Specifically, factors such as knowledge levels, attitudes, perceived risks, and individual differences have a positive correlation with individuals' awareness of data protection measures.
- **Hypothesis 2**: A game theoretical approach can provide valuable insights into the psychological factors influencing individuals' awareness of data protection measures. By analyzing player strategies, moves, rewards, and observations within the game, a deeper understanding of these factors can be obtained, leading to more effective data protection awareness campaigns and educational initiatives.

The structure of this paper is as follows: Section 2 provides background information pertaining to games designed for data protection. Section 3 introduces the proposed data protection game, while Section 5 offers a conclusion for this paper.

## 2 Game Theory for Data Protection

Game theory, a branch of mathematics and economics, provides a powerful framework for analyzing strategic decision-making in interactive situations [4,17]. Traditionally applied in fields such as economics, political science, and biology, game theory offers a unique perspective on human behavior, considering the interplay of individual choices, incentives, and outcomes within a competitive or cooperative setting [4,18]. In recent years, researchers have recognized the potential of game theory in addressing complex issues related to data protection [18].

By incorporating game theoretical elements into the study of data protection, valuable insights can be gained into the psychological factors that influence individuals' awareness and understanding of data protection measures. This research seeks to explore the application of game theory in the context of data protection and investigate the psychological factors influencing individuals' awareness of data protection measures. The research delves into the intersection of game theory and data protection, aiming to uncover the underlying psychological factors that influence individuals' behaviors and attitudes towards safeguarding their data.

### 2.1 Game theory for cyber defense

Yinuo Du et al. [19] discuss the need for adaptive strategies in addressing cyber threats. They highlight that existing game-theoretic frameworks analyzing cyber deception often overlook the defender and attacker's ability to adapt to real-time observations. To address this gap, their paper introduces an Adaptive Cyber Deception Game, a two-player Markov game model [20] that incorporates sequential moves between the defender and attacker [4] in a cyber deception scenario using an attack graph. Additionally, the paper explores the application of Proximal Policy Optimization (PPO), a reinforcement learning algorithm [21] with self-play, within the proposed game model. The objective was to assess the effectiveness of PPO in discovering defender policies that outperform heuristic policies [22].

Preliminary experimental results indicate that the defender policies derived using PPO exhibit significantly better performance compared to heuristic policies [21, 22]. The paper does not extensively examine the psychological factors that impact the behavior and decision-making of defenders and attackers in data protection games, as depicted in Fig. 2. This omission limits the understanding of how these factors affect the effectiveness and outcomes of the games [4,19–22]. The hypergame theory was used in [23] to allow the analysis of conflicts arising from differing viewpoints among multiple players in the context of data protection games.

Hyper game theory considers each player's subjective beliefs, misbeliefs, and perceived uncertainty, thereby influencing their decision-making process in selecting the optimal strategy [4, 23]. By employing hypergame theory, Wan Z et al. [23] aims to provide a robust decision-making mechanism in situations where players hold varying beliefs regarding data protection.

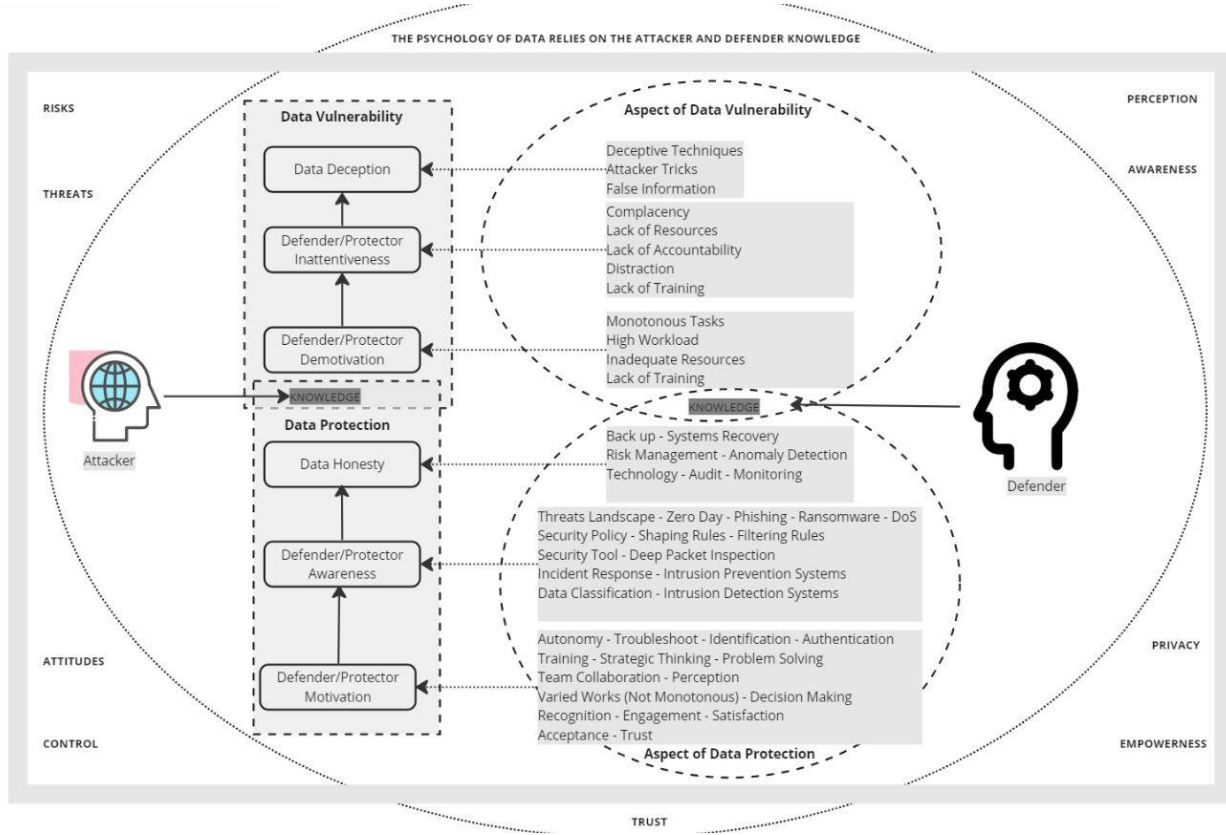

**Fig.2.** Attacker and defender psychological factors in data protection.

Nevertheless, there are limitations to consider when studying psychological factors of data protection within the framework of hypergame theory. Firstly, the subjective beliefs and misbeliefs of players may introduce bias and uncertainty into the decision-making process. These subjective factors can differ significantly among players, making it challenging to establish a standardized approach to data protection. Additionally, the application of hypergame theory assumes rational decision-making by the players.

However, human behavior is influenced by various cognitive and emotional factors that may deviate from strict rationality [2]. Psychological factors such as biases, emotions, and social influences can significantly impact decision-making in data protection scenarios, and their incorporation into the hypergame model may require further exploration. Furthermore, hypergame theory may not fully capture the complexity of psychological factors involved in data protection games. It primarily focuses on players' beliefs and decision-making strategies [23], neglecting deeper psychological aspects such as motivation, attitudes, and per-ception of risks and privacy [2]. These factors play a crucial role in shaping individuals' behaviors and actions related to data protection, and their exclusion from the hypergame framework may limit the comprehensive understanding of psychological dynamics within data protection games. To address these limitations, future research should consider incorporating a more comprehensive understanding of psychological factors into the hypergame model.

This could involve integrating psychological theories and concepts related to risk perception, motivation, and decision-making processes [2]. By doing so, a more nuanced and holistic understanding of how psychological factors influence data protection strategies and outcomes can be achieved. Garg and Grosu [24] established a game-theoretic deception framework and investigated the mixed strategy equilibrium [4] by exploiting deception in attacker-defender interactions in a signaling game with perfect and hybrid Bayesian equilibrium [4]. They thought about defensive deception strategies including honeypots, camouflaged systems, and normal systems.

In contrast to the studies [4, 23–25] that have limited consideration for the psychological aspects of human decision-making and perception, our research considers crucial psychological factors such as cognitive biases, emotions, and social influences [2]. These factors have a profound impact on human behavior in real-world scenarios. By incorporating these psychological factors into the game-theoretic framework for data protection, we aim to provide a more comprehensive understanding of the dynamics involved.

Cognitive biases can lead individuals to deviate from optimal data protection strategies. Emotions can impact risk perception and decision- making, while social influences can shape behavior through peer pressure or conformity [2]. By considering these psychological factors, our research seeks to enhance the effectiveness of data protection methods and techniques by accounting for the complexities of human behavior such as emotional factors, user experience, and cognitive load. This broader perspective can help identify potential limitations in existing approaches and inspire the development of more robust and adaptive strategies in the field of data protection or cybersecurity [26–32].

## 3   The Data Protection Awareness Game

This section presents a novel game theoretical approach to assess data protection awareness. The game pits two players, an attacker and a defender, against each other in a competition [4]. The defender aims to increase their cybersecurity awareness score by strategically placing defensive images on the game board, while the attacker seeks to maximize their intrusion score by placing images that represent attacks. The game is designed to simulate real-world scenarios and challenge players to think ahead and consider all possible outcomes. The goal of the game is to provide better visualization of attacking and defensive patterns to promote data protection awareness. Analyzing the strategies used by both players enables to gain a better understanding of common attack and defense strategies, which can inform training and education efforts and result in a more secure and conscious workforce. Overall, the game theory approach presented in this section has the potential to enhance data protection awareness and promote a more secure work environment. This section highlights the importance of game theory in cybersecurity and presents a practical and engaging tool for assessing and improving data protection awareness.

### 3.1 Knowledge representation and reasoning

The game is divided into three phases.

**Phase 1: Initialization**

In Phase 1, also known as the initialization phase [4], each player is given a set of 13 tokens or images that represent attacking or defending strategies (Table 1).

**Table 1.** The attacker and defender tokens

| Token | Image | Definition | Attacker Trick |
|---|---|---|---|
| A1 | Email | Malicious e-mail | Deceptive |
| A2 | Phone | Malicious phone call | False information |
| A3 | Chat | Malicious chat | Threats |
| A4 | Attachment | Malicious attachment | Deceptive |
| A5 | Donate | Malicious directory | False information |
| A6 | Password | Malicious directory | Lack of training |
| A7 | Connection | Malicious network connection | Distraction |
| A8 | Access | Malicious intrusion | Lack of accountability |
| A9 | Data | Malicious data | Lack of technology |
| A10 | Data loss | Any data loss process | Lack of accountability |
| A11 | Click | Malicious link | False information |
| A12 | Sensitive data | Theft of data | Lack of training |
| A13 | Message | Malicious communication | Threat |
|  |  |  | **Defender Trick** |
| D1 | Denying | Blocking and denying action | Risk Management |
| D2 | Network monitoring | Network traffic analysis | Audit |
| D3 | Avoid clicking | Refuse to click | Security policy |
| D4 | Identification | Verification | Strategic thinking |
| D5 | No trust | Zero trust policy | Intrusion prevention |
| D6 | Upload | Uploading process | Training |
| D7 | Trust | The defender trusts | Risk management |
| D8 | Provide | Providing information | Autonomy |
| D9 | Confidential | Sensitive data | Data classification |
| D10 | Report | Reporting cyber incidents | Treats landscape |
| D11 | Social media | Sharing information | Collaboration |
| D12 | Connection | Trusted network connection | Security tool |
| D13 | Backup | Data recovery | Incident response |

The attacker's tokens are labelled with ($A1, A2, A3, ..., A13$), and the defender's tokens are labelled with ($D1, D2, D3, ..., D13$) (Table 1). During the game, each player takes turns placing their tokens on the game board, with each token placement representing a move that determines whether the attacker or defender used an optimal strategy [4]. The game starts with an empty board, and the judging entity assigns a positive reward of 1 to the player who makes the best move by placing the correct token on the board intersection. A negative reward of 0 is assigned to the player who makes a wrong move [33].

The game format is designed to simulate real-world cybersecurity scenarios and help a person gain a deeper understanding of data security best practices. The game's clear rules and sequential nature make it easy for players to understand and engage with, while the judging entity provides motivation for players to perform their best. The strategies of the game require players to think strategically about their moves and anticipate their opponent's next move. Successful players need to balance their offensive and defensive moves to gain an advantage on the board [4, 33].

The objective of the game is to simulate a scenario in which the defender and attacker compete to accomplish their respective objectives. The defender aims to obtain the highest cybersecurity awareness score by either correctly placing the appropriate image on the board or using the most efficient cybersecurity strategy to defend against the attacker's actions. Conversely, the attacker strives to achieve the highest intrusion score by implementing the most effective attacking strategy to penetrate the defender's cybersecurity defense.

Winning the game necessitates that the defender understands the various cybernetic strategies that the attacker may utilize and can select the optimal defense approach to combat each one. The defender must be watchful and put the correct image on the board to thwart the attacker's successful attacks. In the same vein, the attacker must leverage their knowledge of data protection vulnerabilities to score points. To counter such attacks, it is essential to have a comprehensive data protection strategy that includes both technical and non-technical measures [8, 34, 35].

**Phase 2: Strategy and optimal moves**

During Phase 2 of the game, players are required to create the best possible strategy by moving the image in the game circle (Fig. 3). The move allows players to refine their strategies based on the movements of their opponent. The game is divided into three circles or levels, and tokens can only be placed in a clockwise order (Fig. 3 and Fig. 4). To compute the final score of each player when the game is over, the judging entity evaluates the tokens as illustrated in Fig. 3.

**Phase 3: The scoring schemes**

In Phase 3, the game comes to an end when a player has no moves left or when all 13 images have been placed on the board [4].

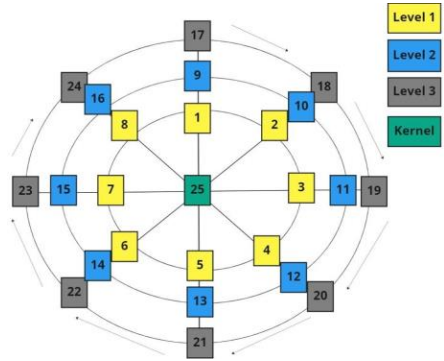

**Fig.3.** The game consists of 25 intersections where tokens can be placed. First round evaluation: [25, 1], [9, 17], [25, 5], and [13, 21]. Second round: [25, 3], [11, 19], [25, 7], and [15, 23]. Third round: [25, 2], [10, 18], [25, 6], and [14, 22]. Fourth round: [25, 8], [16, 24], [25, 4], and [12, 20].

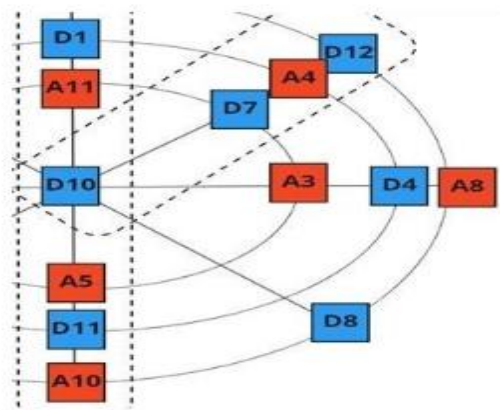

**Fig.4.** Sequential tokens for D10 and D7.

If both players have the same total score, the game is drawn [33]. The judging entity assigns a score ranging from zero to two, depending on the player's performance/moves. A score of zero is given when no optimal move or strategy is used, a score of one is given when the player makes the best move by placing the correct image, and a score of two is given when the player places two images sequentially without the opponent's obstruction. Moreover, it is important to note that a player cannot have two of their tokens placed sequentially on the board, as this would result in a score of two being assigned to that player (see Fig. 4 with D10 and D7). In the context of a data protection game, the judging entity can analyze the implications of the defender strategies shown in Fig. 4 (D10 and D7) as follows:

1. Defender Strategy D10 (Reporting Cybercrime): This strategy involves the defender taking a proactive approach to combat cybercrime by promptly reporting any suspicious or unlawful activities to the relevant authorities [8]. By reporting cybercrimes, the defender aims to contribute to the overall security and protection of data [17]. This strategy indicates a strong commitment to data protection and a willingness to take action against potential threats.
2. Defender Strategy D7 (Trusting Risk Management): This strategy reflects the defender's reliance on effective risk management practices to ensure data protection [12]. By trusting risk management, the defender acknowledges the importance of implementing preventive measures, risk assessments, and security protocols to mitigate potential threats [21]. This strategy suggests a belief in the effectiveness of risk management processes and the ability to make informed decisions based on risk evaluations.

This scoring mechanism implies that the game values and rewards defenders who prioritize both reporting cybercrime and trusting risk management. The significance of these strategies in a data protection game lies in their contribution to maintaining data security and safeguarding against potential breaches or cyber-attacks [26].

By reporting cybercrime, the defender helps in the identification and prosecution of perpetrators, aiding in the prevention of future incidents. Trusting risk management ensures the implementation of robust security measures, reducing the likelihood of data breaches and ensuring compliance with data protection regulations [8, 28]. In terms of psychological factors, these strategies can be linked to the following:

1. Sense of Responsibility: Both strategies demonstrate a sense of responsibility towards data protection. The defender recognizes the importance of taking proactive measures to safeguard data and actively contribute to the overall security ecosystem [36].
2. Trust and Confidence: The defender's trust in risk management indicates confidence in the effectiveness of preventive measures and security protocols. This mindset fosters trust in the overall data protection infrastructure and encourages a proactive approach to mitigating risks.
3. Alignment with Ethical Values: The adoption of these strategies aligns with ethical principles of data protection and integrity [5]. The defender prioritizes transparency, accountability, and integrity by reporting cybercrime and relying on risk management to protect sensitive data.
4. Compliance and Regulatory Awareness: These strategies reflect the defender's awareness and adherence to data protection regulations. By reporting cybercrime and trusting risk management, the defender demonstrates a commitment to complying with legal requirements and maintaining a secure data environment.

In summary, the defender strategies of reporting cybercrime (D10) and trusting risk management (D7) play a vital role in a data protection game [15,25].

The scoring scheme used by the judging entity has been illustrated in Table 2 where A% and D% represent the attacker and defender score. It is important to note that the same token can only be used twice on the board. Possible attacker images are depicted in Fig. 5. However, the types and number of images can be expanded.

**Table 2.** An instance of the scoring table

| Iteration | A | D | A% | D% | Judge | Comments |
|---|---|---|---|---|---|---|
| 1 | Email | Zero trust | 0 | 1 | Defender best move | Never trust malicious emails |
| 2 | Click | Denying | 0 | 1 | Defender best move | Denied malicious link |
| 3 | Chat | Identification | 0 | 1 | Defender best move | Identified malicious chats |
| 4 | Phone | Trust | 1 | 0 | Attacker best move | Malicious calls trusted |
| 5 | Connection | Connection | 0 | 1 | Defender best move | Secure connections suggested |
| 6 | Access | Identification | 0 | 1 | Defender best move | Data access monitored |
| 7 | Data loss | Backups | 0 | 1 | Defender best move | The defender data recovery |
| 8 | Message | Identification | 0 | 1 | Defender best move | Abnormal message identified |
| 9 | Click | Upload | 1 | 0 | Attacker best move | The defender uploaded files |
| 10 | Password | Provide | 1 | 0 | Attacker best move | The defender shared passwords |
| 11 | Data | Network monitoring | 0 | 1 | Defender best move | Malicious data monitored |
| 12 | Donate | No trust | 0 | 1 | Defender best move | The defender did not share data |
| 13 | Donate | Provide | 1 | 0 | Attacker best move | Device validation details shared |
| 14 | Donate | Social media | 1 | 0 | Attacker best move | Relevant information shared |
| 15 | Connection | Report | 0 | 1 | Defender best move | Malicious connection reported |
| 16 | Access | Connect | 0 | 1 | Defender best move | Secure connection |
| 17 | Data loss | Provide | 1 | 0 | Attacker best's move | The defender lost information |
| 18 | Click | Identification | 0 | 1 | Defender best's move | Malicious link identification |
| 19 | Message | Backups | 1 | 0 | Attacker best's move | The defender shared backups |
| 20 | Attachment | Avoid | 0 | 1 | Defender best's move | Malicious attachments avoided |
| 21 | Chat | Trust | 1 | 0 | Attacker's best move | Malicious chat trusted |
| 22 | Phone | Network monitoring | 0 | 1 | Defender best's move | Secure network monitored |
| 23 | Data | Avoid | 0 | 1 | Defender best's move | Abnormal data avoided |
| 24 | Sensitive data | Backup | 0 | 1 | Defender best's move | Data recovery |
| 25 | Password | Avoid | 0 | 1 | Defender best's move | Secured passwords |
| 26 | Data loss | Upload | 1 | 0 | Attacker best's move | Information uploaded and lost |

### 3.2   Limitation and future works

The study lacks specific information about the participants' sample size and characteristics, making it challenging to determine the findings' generalizability to a broader population. Additionally, there could be other relevant factors that were not considered in this study regarding individuals' engagement with data protection practices.

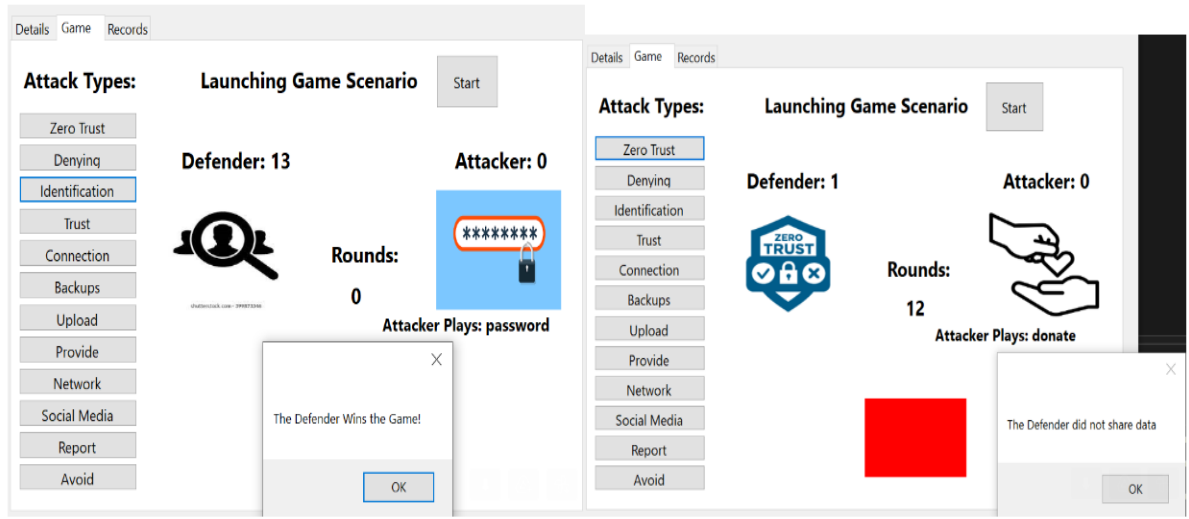

**Fig.5.** The attacker images.

To address this, future research directions can include a comparative analysis of individuals from diverse demographic backgrounds (e.g., age, gender, education level) to examine potential significant differences in their awareness and understanding of data protection measures, thereby enhancing the findings' generalizability. A longitudinal study can be conducted to explore how individuals' awareness and understanding of data protection evolve over time, providing insights into the dynamics of data protection attitudes and behaviors, especially in response to changing technologies and societal factors.

Furthermore, implementing the game with a focus on assessing the usability and effectiveness of various strategies and interfaces can help identify barriers and challenges individuals face in engaging with data protection practices, leading to the development of more user-friendly solutions.

Conducting a cross-cultural analysis using the game can offer insights into how cultural values, norms, and beliefs shape individuals' attitudes and behaviors related to data protection, enabling a better understanding of the cultural influences involved. Lastly, studies can be conducted to design and implement educational initiatives or interventions aimed at enhancing individuals' awareness and understanding of data protection measures. Evaluating the effectiveness of these interventions will contribute to a more comprehensive understanding of the psychological factors influencing individuals' engagement with data protection practices. Addressing these research directions would help overcome the limitations of the current study, providing valuable insights into the psychological factors impacting individuals' engagement with data protection practices.

## 4   Conclusion

This article introduces a unique method that utilizes game theory to explore the psychological factors impacting data protection practices. The study uncovers important insights into individuals' attitudes and behaviors towards data protection, such as knowledge levels, attitudes, perceived risks, and individual differences. These findings contribute to an understanding of the complex relationship between psychology and data protection, and they have practical implications for developing effective awareness campaigns and educational initiatives in this area.